%% file: charm2015_ClaudiaVacca.tex
\documentclass[12pt]{article}
\usepackage{graphicx}
\usepackage{subfig}

\newcommand\pubnumber{WSU--HEP--XXYY}
\newcommand\pubdate{\today}

\def\wayne{Department of Physics\\
Universit\'a degli Studi di Cagliari, Cagliari, ITA,\\
INFN Sezione di Cagliari,\\
CERN
}
\def\support{\footnote{on behalf of LHCb collaboration.}}

\def\Title#1{\begin{center} {\Large #1 } \end{center}}
\def\Author#1{\begin{center}{ \sc #1} \end{center}}
\def\Address#1{\begin{center}{ \it #1} \end{center}}

\newcommand\pubblock{\rightline{\begin{tabular}{l} \pubnumber\\
         \pubdate  \end{tabular}}}
\newenvironment{Abstract}{\begin{quotation}  }{\end{quotation}}
\newenvironment{Presented}{\begin{quotation} \begin{center} 
             PRESENTED AT\end{center}\bigskip 
      \begin{center}\begin{large}}{\end{large}\end{center} \end{quotation}}

\input econfmacros.tex

\begin{document}
\begin{titlepage}
\pubblock

\vfill
\Title{Measurements of charm rare decays at LHCb}
\vfill
\Author{Claudia Vacca\support}
\Address{\wayne}
\vfill
\begin{Abstract}
Following the intriguing hints of deviations from the Standard Model in rare B meson decays, searches for rare and forbidden decays of charm hadrons become a hot topic again. We present recent results on Flavour Changing Neutral Current $D^{0}\rightarrow\mu^+ \mu^-$, $D^0 \rightarrow\mu^+\mu^- \pi^+ \pi^-$, $D^{\pm}_{(s)}\rightarrow \pi^{\pm} \mu^+\mu^-$ and LFV $D^{\pm}_{(s)}\rightarrow \pi^{\mp}\mu^{\pm}\mu^{\pm}$ obtained at LHCb.

Some future prospects on the field of charm rare decays are also explained as well as some predictions on what is expected after LHC Run II and LHCb Upgrade.  
\end{Abstract}
\vfill
\begin{Presented}
The 7th International Workshop on Charm Physics (CHARM 2015)\\
Detroit, MI, 18-22 May, 2015
\end{Presented}
\vfill
\end{titlepage}
\def\thefootnote{\fnsymbol{footnote}}
\setcounter{footnote}{0}
%

\section{Introduction}

Charm rare decays represent a unique chance to investigate Flavour Changing Neutral Currents (FCNC) processes mediated by up-type quarks. These studies, which are complemetary to those on B and K sectors, together with those on Lepton Number Violating (LNV) processes, could allow to test Standard Model predictions and have a privileged point of view on confirming or rejecting New Physics (NP) theories.     
In fact, FCNC processes are highly suppressed on Standard Model and only allowed at loop level and the Glashow-Iliopoulos-Maiani (GIM) suppression affects D decays, involving $c\rightarrow u \mu^+\mu^-$ processes, more than B ones, due to the absence of a high-mass down-type quark.

The extremely low branching fraction predicted by the SM make the $D\rightarrow h(h)\mu\mu$ decays be perfect candidates to investigate NP models predicting enhancements up to several orders of magnitude. 

Furthermore, multibody semileptonic decays  allow angular studies such as  searches for Forward-backward asymmetries, which could also be enhanced by some NP effects to $\mathcal{O}(1\%)$ \cite{10} \cite{11}.

\section{The LHCb detector}

The Large Hadron Collider-beauty (LHCb) detector is a single-arm forward spectrometer, aimed to study b- and c-hadrons rare decays, CP violation, test the quark model and investigate the physics beyond the Standard Model (matter-antimatter
asymmetry).
For this aim, it works under low luminosity conditions, with few p-p interactions per bunch crossing to ensure a better
reconstructibility of events.
These properties, together with an excellent muon identification, a high momentum resolution ($0.4\% <
\delta p/p < 0.6\%$), very good performance in reconstruction of vertices and a high performance trigger, flexible and configurable make LHCb be a very suitable detector for studying charm rare decays.
Moreover, $b\bar{b}$/$c\bar{c}$ pairs are predominately produced at high $|\eta|$ and 5(2)$\cdot 10^{12}$ $D^{0}(D^{+})$ have been produced in LHCb acceptance (1.9 $< \eta < $4.9) in 3$fb^{-1}$ of integrated luminosity at $\sqrt{s}$ = 7-8 TeV.

\section{Search for the rare decay $D^{0}\rightarrow \mu^{+} \mu^{-}$}
The measurement on $D^0\rightarrow \mu^+ \mu^-$ branching fraction \cite{2} has been performed by analysing a data sample corresponding to an integrated luminosity of 0.9 $fb^{-1}$ of pp collisions collected at a centre-of-mass energy of 7 TeV by the LHCb experiment.

This FCNC process is expected to be extremely rare in SM, due to an additional helicity suppression. The short distance contributions ($\mathcal{O}(10^{-18})$) \cite{22} are negligible with respect to long distance ones which are dominated by a two-photons intermediate state. The current upper limit on $\mathcal{B}(D^0 \rightarrow \gamma	\gamma)$ is $2.6\cdot 10^{-6}$ \cite{23} and translates into an upper bound for SM predictions of $\mathcal{O}(10^{-11})$.

The previous limit on this branching fraction was set by the Belle Collaboration to $1.4\cdot10^{-7}$ \cite{1}.

The search for the decay is performed using $D^{*+}\rightarrow D^0(\mu^+\mu^−)\pi^+$ decays, with the $D^{*+}$ produced directly at a p-p collision primary vertex.

The $D^{*+}\rightarrow D^0(\pi^+ \pi^-)\pi^+$ has been selected as a normalisation mode and the branching fraction has been obtained from:
\begin{center}
$\mathcal{B}(D^0\rightarrow \mu^+\mu^-) = \frac{N_{\mu^+ \mu^-}}{N_{\pi^+\pi^-}} \cdot \frac{\epsilon_{\pi\pi}}{\epsilon_{\mu\mu}}\cdot \mathcal{B}(D^0 \rightarrow \pi^+\pi^-)
$.
\end{center}

Some control channels ($D^{*+}\rightarrow D^0(K^- \pi^+) \pi^+$, $D^0\rightarrow K^- \pi^+$ without the $D^*+$ assumption, $J/\Psi \rightarrow \mu^+ \mu^-$) have been used to determine muon identification and trigger efficiencies and in particular the $D^0 \rightarrow K^-\pi^+$ sample has been used to control the misidentification rate of the pion as muon, which was one of the major aspects of the analysis.

\begin{figure}[htb]
\centering
\includegraphics[height=2.7in]{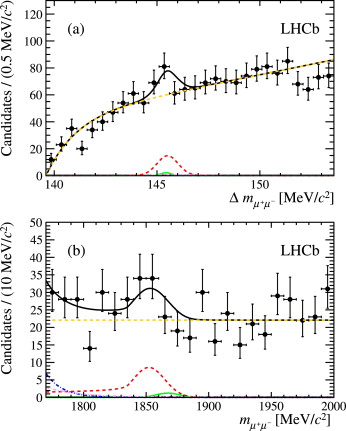}
\caption{\footnotesize{(a) Invariant mass difference with $m_{\mu^+\mu^−}$ in the range 1820-1885 MeV/$c^2$ and (b) invariant mass $m_{\mu^+\mu^-}$, with $\Delta m_{\mu^+\mu^-}=m_{\pi^+\mu^+\mu^-}$ in the range 144-147 MeV/$c^2$ for $D^{*+}\rightarrow D^0(\mu^+\mu^-)\pi^+$ candidates. The projections of the two-dimensional unbinned extended maximum likelihood fit are overlaid. The curves represent the total distribution (solid black), the $D^{*+}\rightarrow D^0(\pi^+ \pi^-)\pi^+$ (dashed red), the combinatorial background (dashed yellow), the
$D^{*+}\rightarrow D^0(K^- \pi^+)\pi^+$  (dash-dotted blue), the $D^{*+}\rightarrow D^0(\pi^- \mu^+ \nu_{\mu})\pi^+$ (dash-dotted purple) and the signal $D^{*+}\rightarrow D^0(\mu^+\mu^-)\pi^+$  (solid green) contribution.}}
\label{fig:mumu1}
\end{figure}

The backgroud originates from two sources:
\begin{itemize}
\item peaking background (2- or 3-body $D^0$ decays in which hadrons are misidentified as muons), reduced by applying tight particle identification criteria;
\item combinatorial background, due to semileptonic decays of beauty and
charm hadrons, suppressed by a multivariate selection based on a boosted decision tree whose input variables are: the $D^0$ pointing angle $\theta_D$, $\chi^2_{IP}$ of $D^0$ vertex and the two muons, minimum $p_T$ of the two muons, $\chi^2_{KF}$ (Kalman filter) of the constrained fit, positively-charged muon angle in the $D^0$ rest frame with respect to the $D^0$ flight direction and $D^0$ angle in the $D^{*+}$ rest frame with respect to the $D^{*+}$ flight direction.

\end{itemize}

\begin{figure}[htb]
\centering
\includegraphics[height=1.7in]{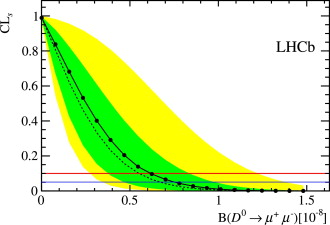}
\caption{\footnotesize{Confidence levels (CLs) (solid line) as a function of the assumed $D^0\rightarrow\mu^+ \mu^−$ branching fraction and
median (dashed line), 1$\sigma$ and 2$\sigma$ bands of the expected CLs, in the background-only hypothesis,
obtained with the asymptotic CLs method. The horizontal line corresponding to CLs=0.05 is also drawn.}}
\label{fig:mumu2}
\end{figure}

The result obtained for the upper limit on this branching fraction is:

\begin{center}
$\mathcal{B}(D^0\rightarrow \mu^+\mu^-) < 6.2(7.6) \cdot 10^{-9}$ at 90$\%$ (95$\%$) CL

\end{center}

and represents an improvement by a factor 20 with respect to the previous limit, but still lets 2 orders of magnitude to be investigated before reaching the value predicted in the SM.

\section{Search for $D^0\rightarrow \pi^+ \pi^- \mu^+ \mu^-$ decay}

 The $D^0\rightarrow \pi^+ \pi^- \mu^+ \mu^-$ decay branching fraction is dominated by long distance
contributions such as $D^0\rightarrow V(\mu^+\mu^-)\pi^+ \pi^-$ $(O(10^{-6}))$, in which muons come from a resonance
($\rho,\phi,\eta$). The measurement performed at LHCb \cite{4} aims at constraining the short distance contribution to this decay. 
This is an FCNC process whose branching fraction is expected to be lower than $10^{-9}$ in the SM and for this reason it can be used to probe NP.

The best limit before this result was $3.0\cdot 10^{-5}$ at 90$\%$ CL (confidence level) and was performed by E791 Collaboration \cite{3}.

The data sample used corresponds to an integrated luminosity of 1.0 $fb^{-1}$ at $\sqrt{s}$ = 7 TeV collected by the LHCb detector.

The sample of $D^0$s is originated from the decay $D^{*+}\rightarrow D^0 \pi^+$.The analysis is performed in di-muon mass regions, to minimize the leakage coming from resonances. The so-called signal regions are defined as $250<m_{\mu\mu}<525MeV/c^2$ and $m_{\mu\mu}>1100MeV/c^2$.

The selection has been performed by combining a multivariate analysis based on BDT(input variables:$\theta_D$, $\chi^2$ of $D^0$ decay vertex and flight distance, $\chi^2_{IP}$, p and $p_T$ of all final state tracks, $\chi^2$ of the vertex and $p_T$ of the $D^{*+}$ candidate, maximum distance of closest approach between all pairs of tracks forming the $D^0$ and $D^{*+}$ candidates) and some requirements on muon particle identification.
A peaking background is originated from the misidentification of pions in the decay $D^0\rightarrow \pi^+ \pi^- \pi^+ \pi^-$, while a secondary source is combinatorial one.

The measurement has been normalised to a reference sample of  $D^0\rightarrow \pi^+\pi^-\phi(\mu^+\mu^-)$ decays. Unfortunately, the branching fraction of this decay has not been measured directly.  Its value has been derived from the fit fractions of the various $\phi(K^+ K^-) \pi^+ \pi^-$ contributions 
found by the amplitudes analysis for the $D^0 \rightarrow K^+ K^- \pi^+ \pi^-$ decay reported in \cite{12}, and from the 
 $\frac{\mathcal{B}(\phi\rightarrow \mu^+ \mu^-)}{\mathcal{B}(\phi \rightarrow K^+ K^-)}$ ratio.
 This procedure results in a systematic
uncertainty of 17$\%$ on the branching fraction of the normalisation mode, which dominated the 
total uncertainty: $\mathcal(B)(D^0\rightarrow \pi^+\pi^- \phi(\mu^+\mu^-)) = (5.2 \pm 1.1)\cdot 10^{-7}$.

The $D^0 \rightarrow \pi^+\pi^-\mu^+\mu^-$ yields have also been fitted in the regions where the $\phi$ and $\rho$ resonances peak. In order to evaluate the size of the 
leakage from resonances into the signal regions, the fraction of such decays lying in these regions was estimated
assuming Breit-Wigner lineshapes. No interference between amplitutes has been considered, since the corresponding effects are
negligible at the level of precision reachable.

\begin{figure}[htb]
\centering
\subfloat[]{\label{main:1}\includegraphics[height=2.8in,angle=90]{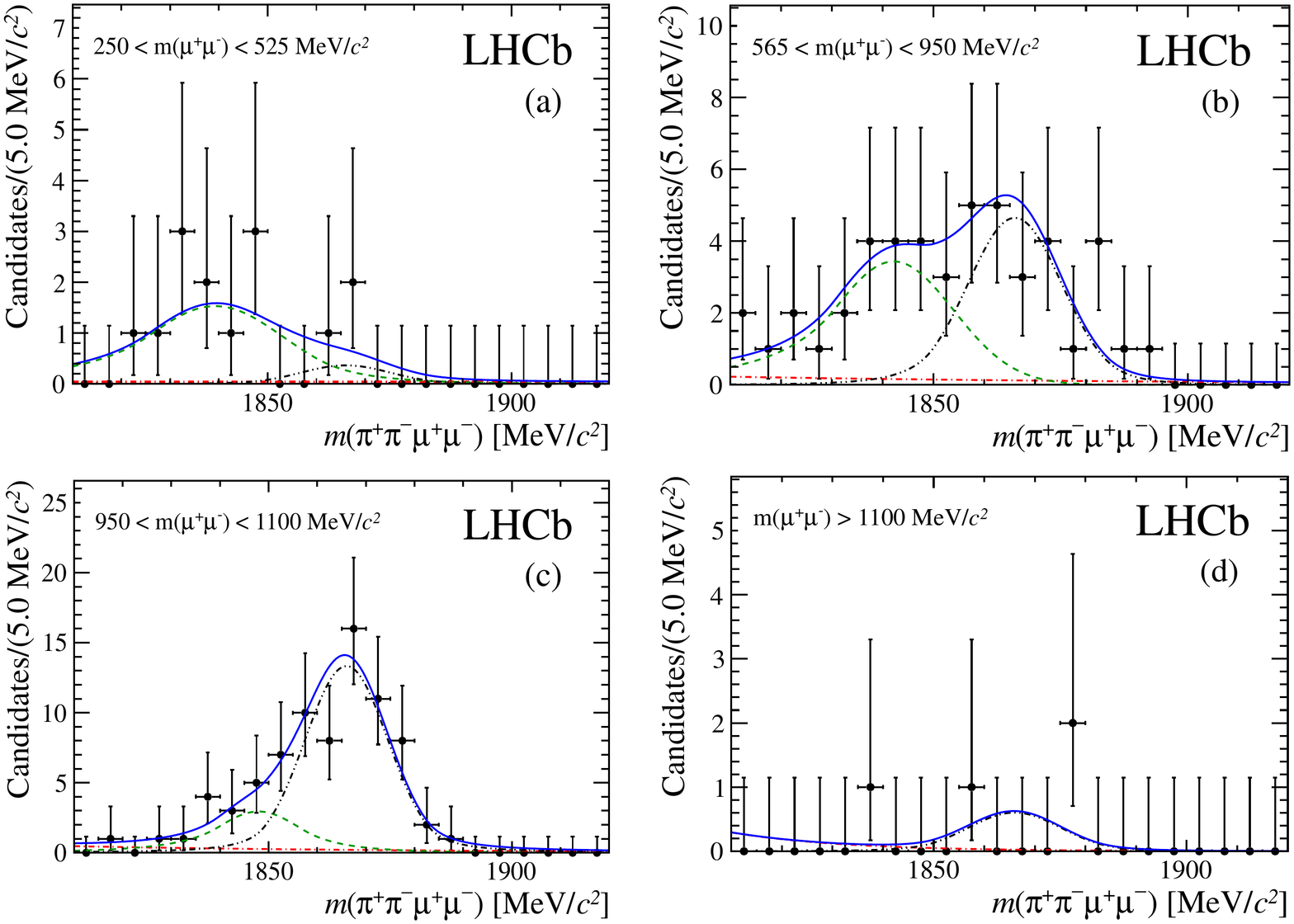}}
\subfloat[]{\label{main:2}\includegraphics[height=2.8in,angle=90]{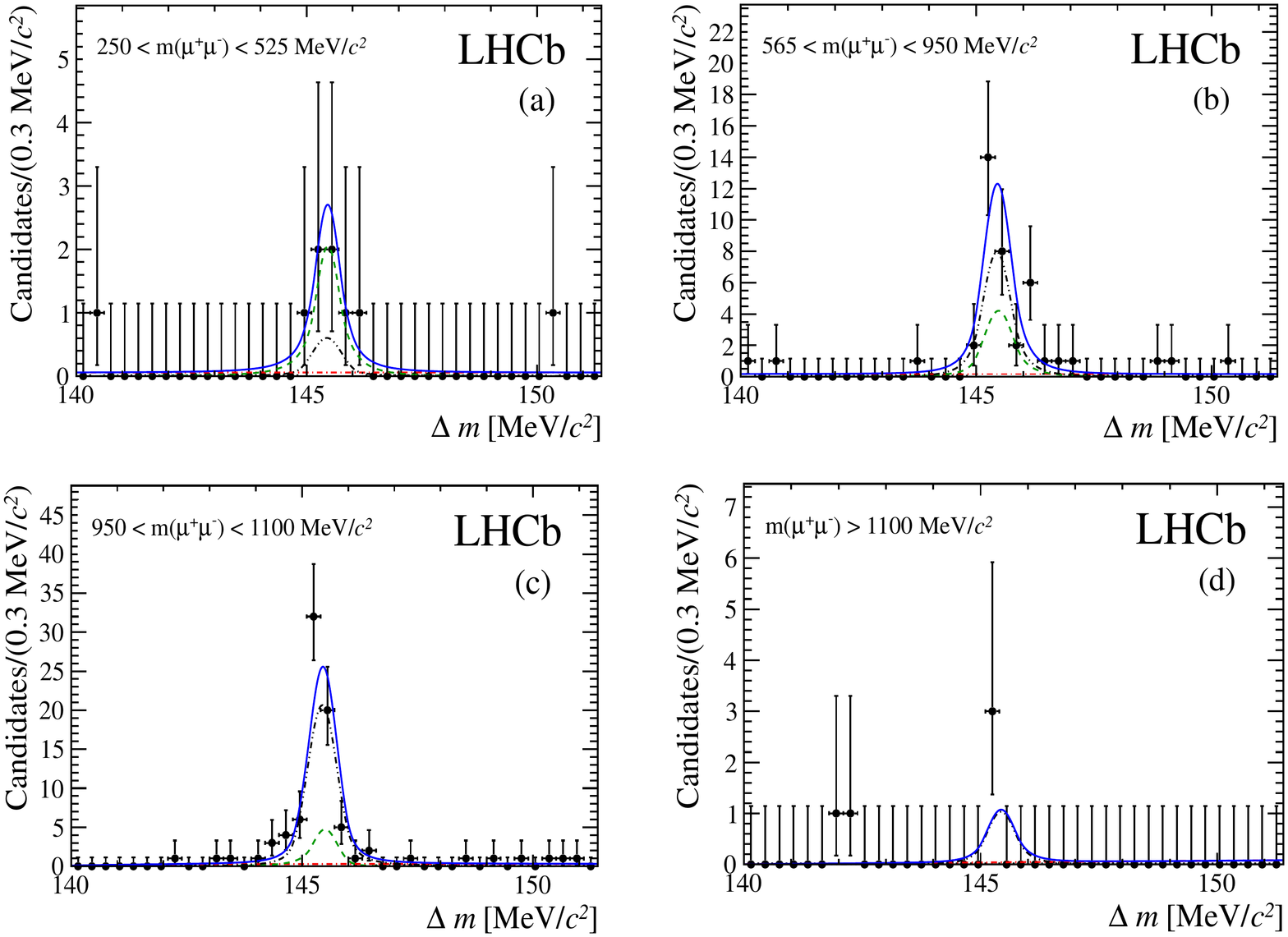}}
\caption{\footnotesize{[1] Distribution of m($\pi^+\pi^-\mu^+\mu^-$) for $D^0\rightarrow \pi^+\pi^-\mu^+\mu^-$ candidates in the a) low m($\mu^+\mu^-$), b)$\rho/\omega$, (c) $\phi$, and (d) high-m($\mu^+\mu^-$) regions, with $\Delta$m in the range 144.4-146.6
MeV/$c^2$.
[2] Distribution of $\Delta$m for $D^0\rightarrow \pi^+\pi^-\mu^+\mu^-$ candidates in the a) low m($\mu^+\mu^-$), b)$\rho/\omega$, (c) $\phi$, and (d) high-m($\mu^+\mu^-$) regions, with $D^0$ invariant mass in the range 1840-1888 MeV/$c^2$.
The data are shown as points (black) and the ﬁt result (dark blue line) is overlaid. The
components of the ﬁt are also shown: the signal (ﬁlled area), the $D^0\rightarrow\pi^+\pi^-\pi^+ \pi^-$ background (green dashed line) and the non-peaking background (red dashed-dotted line).}}
\label{fig:pipimumu1}
\end{figure}

The signal and background yields have been measured using an unbinned maximum likelihood fit of two-dimensional [$m_{\pi\pi\mu\mu}$,$\Delta$m]
distributions in ranges 1810-1920 MeV/$c^2$ and 140-151.4 MeV/$c^2$ respectively, with $\Delta$m=$m_{\pi\pi\mu\mu\pi}-m_{\pi\pi\mu\mu}$ (Fig.\ref{fig:pipimumu1}).

The result obtained for the branching fraction in the signal regions has been extrapolated to the whole dimuon mass window by assuming a phase space model.

The upper limit for the branching fraction is set to:

\begin{center}$\mathcal{B}(D^0\rightarrow \pi^+\pi^-\mu^+\mu^-) < 5.5(6.7) \cdot 10^{-7}$ at 90$\%$ (95$\%$) CL
\end{center}

\begin{figure}[htb]
\centering
\includegraphics[height=1.6in]{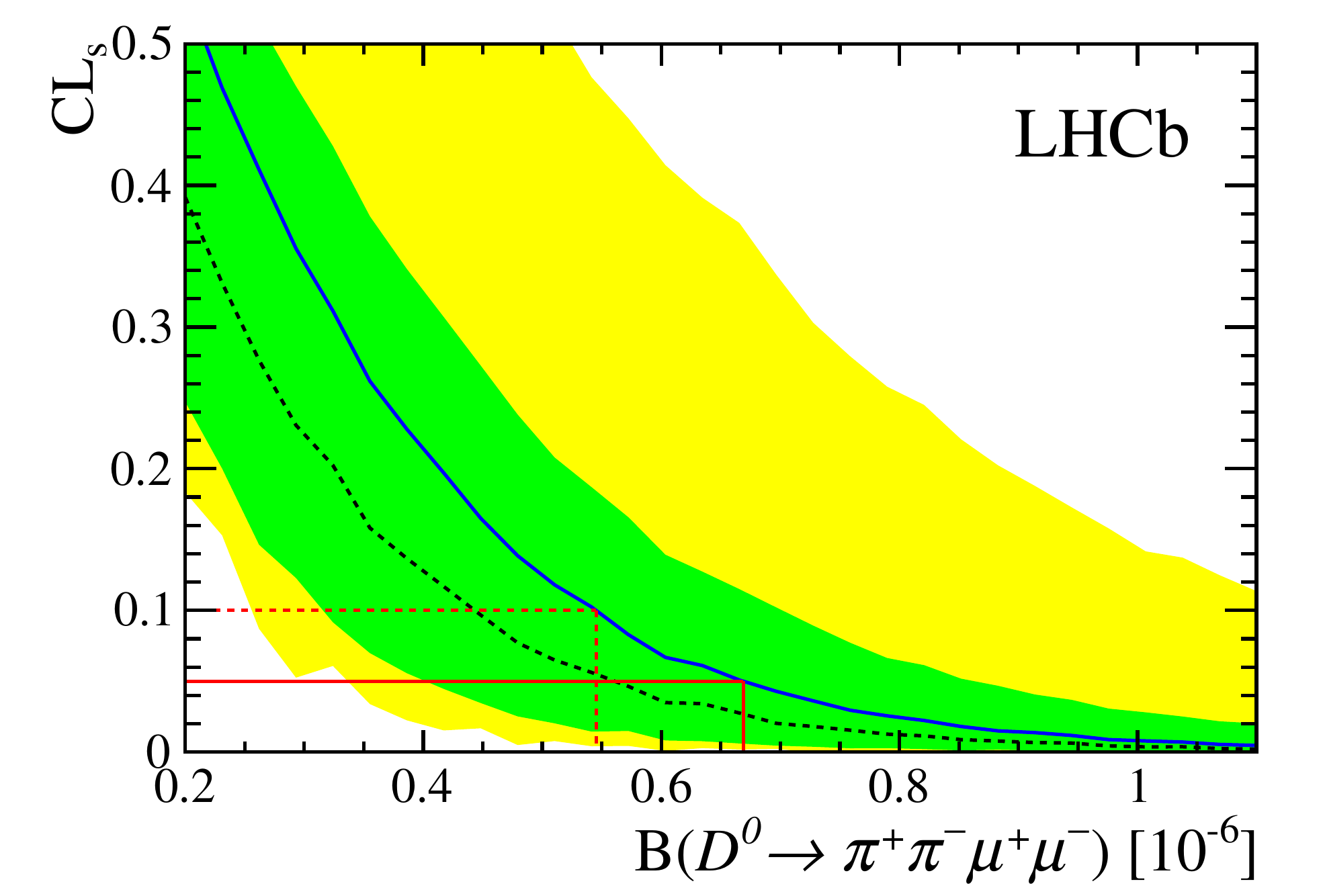}
\caption{\footnotesize{Observed (solid curve) and expected (dashed curve) CLs values as a function of $\mathcal{B}(D^0\rightarrow\pi^
+\pi^-\mu^+\mu^-)$. The green (yellow) shaded area contains 68.3$\%$ and 95.5$\%$ of the results
of the analysis on experiments simulated with no signal. The upper limits at the 90(95)$\%$ CL
are indicated by the dashed (solid) line.}}
\label{fig:pipimumu5}
\end{figure}

This limit improves the previous one by a factor almost 70, but still lays 2 orders of magnitude above SM predictions.

\section{Search for $D_{(s)}^+ \rightarrow \pi^+ \mu^+ \mu^-$ / $D_{(s)}^+ \rightarrow \pi^- \mu^+ \mu^+$ decays}

The measurement of the $D_{(s)}^+ \rightarrow \pi^+ \mu^+ \mu^-$ and $D_{(s)}^+ \rightarrow \pi^- \mu^+ \mu^+$ \cite{5} branching fractions has been performed at LHCb using proton-proton collision data, corresponding to an integrated luminosity of 1.0
$fb^{-1}$ at $\sqrt{s}$=7 TeV recorded in 2011.

While the $D^+ \rightarrow \pi^+ \mu^+ \mu^-$ decay is dominated by $D^+\rightarrow \pi^+ V(\mu^+\mu^-)$
long distance contributions and also receives contribution from the FCNC process,
we aimed at probing that the $D_{s}^+ \rightarrow \pi^+ \mu^+ \mu^-$ proceeds via a weak annihilation CKM suppressed. The latter can also be used to normalise a potential weak annihilation contribute in the $D^+ \rightarrow \pi^+ \mu^+ \mu^-$.
The limits set for these processes before LHCb measurement were 
\begin{center}
$\mathcal{B}(D^+ \rightarrow \pi^+ \mu^+ \mu^-)< 3.9\cdot 10^{-6}\,\cite{7}$ 

$\mathcal{B}(D_{s}^+ \rightarrow \pi^+ \mu^+ \mu^-) <  2.6\cdot 10^{-5} \, \cite{8}$

\end{center}

The $D_{(s)}^+ \rightarrow \pi^- \mu^+ \mu^+$ decay is a LNV (Lepton Number Violation) process, and thus prohibited in the SM, since it could only happen via lepton mixing mediated by non -SM particle as a Majorana neutrino.
The previous limits established for these decays were $\mathcal{O}(10^{-6})$ for the $D^+ \rightarrow \pi^- \mu^+ \mu^+$ \cite{6} and one order of magnitude above for the $D_{s}^+ \rightarrow \pi^- \mu^+ \mu^+$ \cite{6}.

The measurement has been performed using the $D^+_{(s)}\rightarrow \pi^+\phi(\mu^+\mu^-)$ as a control channel. The only peaking background contributing is that originated from $D^+_{(s)}\rightarrow \pi^+ \pi^+\pi^-)$. 
Candidates have been selected via a multivariate analysis based on BDT (the input variables used are $\theta_D$,$\chi^2$ of $D^+_{(s)}$ decay vertex and fligh distance,p and $p_T$ of all tracks, IP $\chi^2$ of all tracks, maximum distance of closest approach between all pairs of tracks in the candidate $D^+_{(s)}$ decay) and some requirements on particle identification variables.

\begin{figure}
\begin{center}
\subfloat[]{\label{main:1}\includegraphics[width=2.8in]{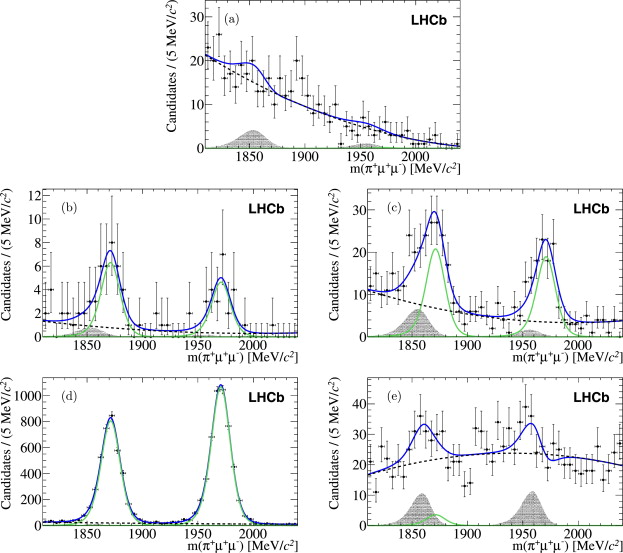}}
\subfloat[]{\label{main:2}\includegraphics[width=2.8in]{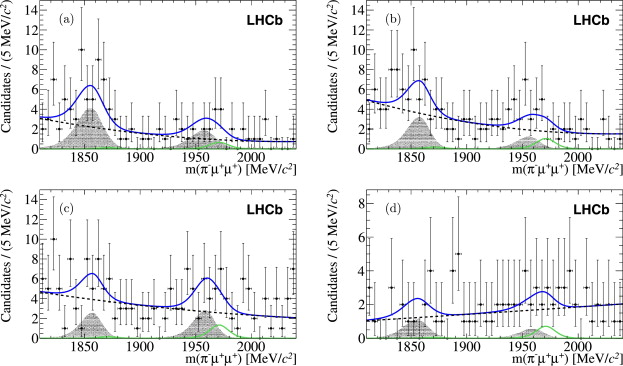}}
\caption{\footnotesize{Invariant mass distributions for: 
[1]$D^+_{(s)}\rightarrow \pi^+ \mu^+\mu^-$ candidates in the ﬁve m($\mu^+\mu^-$) bins: a) low-m($\mu^+\mu^-$) [250-525]$MeV/c^2$, b) $\eta$ [525-565]$MeV/c^2$, c) $\rho/\omega$  [565-850]$MeV/c^2$, d) $\phi$ [850-1250]$MeV/c^2$, e) high-m($\mu^+\mu^-$) [1250-2000]$MeV/c^2$; [2]$D^+_{(s)}\rightarrow \pi^- \mu^+\mu^+$ candidates in the ﬁve m($\mu^+\pi^-$) bins: a) 250-1140 $MeV/c^2$, b) 1140-1340 $MeV/c^2$, c) 1340-1550 $MeV/c^2$, d) 1540-2000 $MeV/c^2$.
\\ The data are shown as points (black)and the total PDF (dark blue line) is overlaid. The components of the ﬁt are also shown: the signal (light green line), the peaking background (solid area) and the non-peaking background (dashed line).}}
\label{fig:pimumu1}
\end{center}
\end{figure}

The yields have been obtained with a binned maximum likelihood fit. Details are shown in Fig.\ref{fig:pimumu1} in m($\mu^+\mu^-$) bins.


\begin{figure}
\begin{center}
\includegraphics[width=2in]{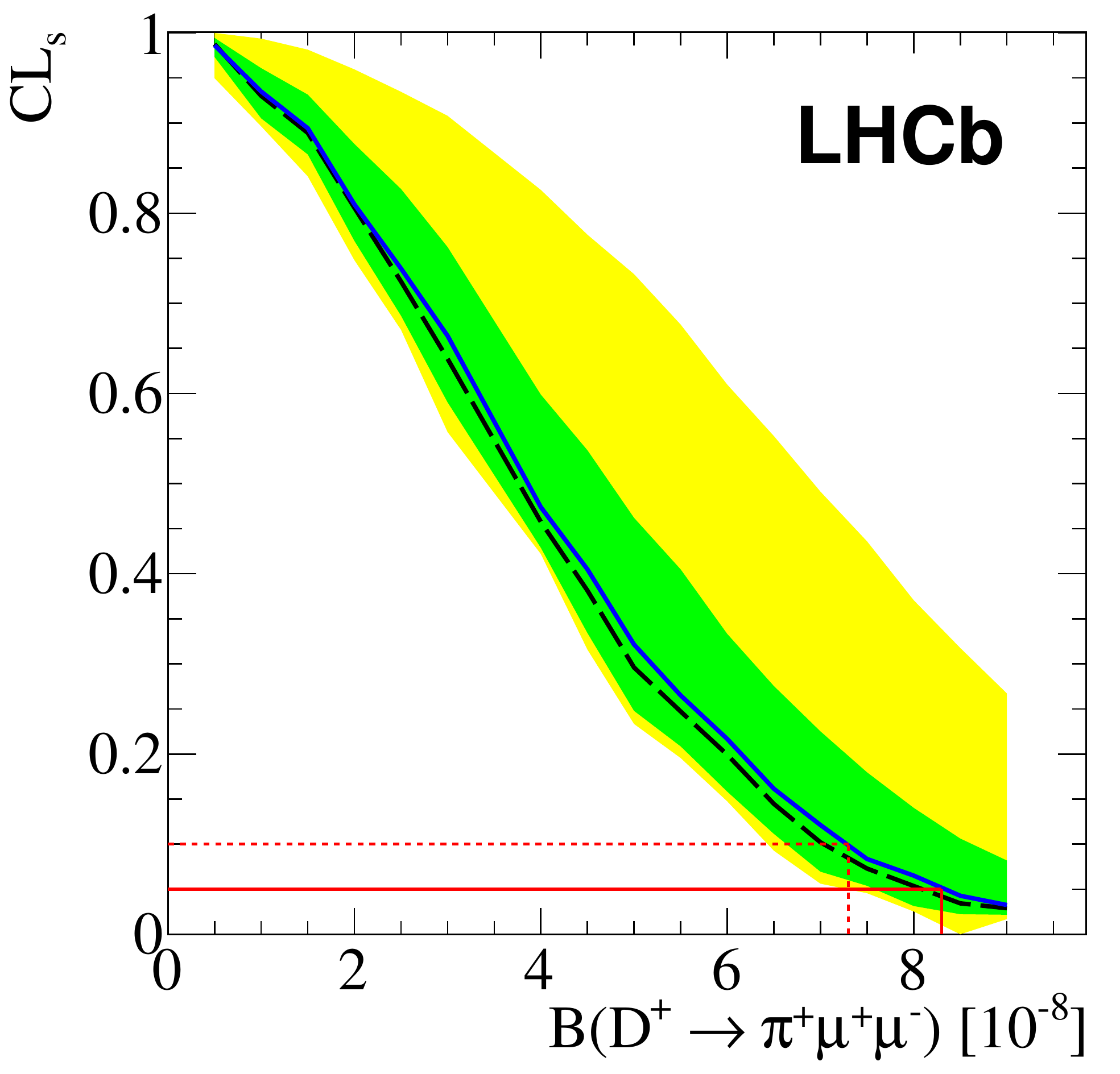}
\caption{\footnotesize{Observed (solid curve) and expected (dashed curve) CLs values as a function of $\mathcal(B)(D^+ \rightarrow \pi^+\mu^+\mu^-$). The green (yellow) shaded area contains the $\pm 1\sigma$ ($\pm2\sigma$) interval of possible
results compatible with the expected value if only background is observed. The upper limits at the 90$\%$ (95$\%$) CL are indicated by the dashed (solid) line.}}
\label{fig:pimumu3}
\end{center}
\end{figure}

The results extrapolated to the full dimuon mass spectrum are:
\begin{center}
$\mathcal{B}(D^+ \rightarrow \pi^+ \mu^+ \mu^-) < 7.3(8.3) \cdot 10^{-8}$ at $90\% (95\%)$ CL

$\mathcal{B}(D_s^+ \rightarrow \pi^+ \mu^+ \mu^-) < 4.1(4.8) \cdot 10^{-7}$ at $90\% (95\%)$ CL

$\mathcal{B}(D^+ \rightarrow \pi^- \mu^+ \mu^+) < 2.2(2.5) \cdot 10^{-8}$ at $90\% (95\%)$ CL

$\mathcal{B}(D_s^+\rightarrow \pi^-\mu^+\mu^+) < 1.2(1.4) \cdot 10^{-7}$ at $90\% (95\%)$ CL
\end{center}

and represent an improvement by a factor 50 with respect to previous limits, but are still one order of magnitude above several NP predictions on  $D_{(s)}^+\rightarrow \pi^+\mu^+\mu^-$ \cite{9}.

\section{Future prospects}

The strictest limit on the results currently obtained on charm rare decays field is certainly the statistic available.

As it is well known, LHC is imminently starting with Run II data taking, which in the specific case of LHCb will enhance the energy in the centre of mass from 8 to 13 TeV. At the end of Run II, LHCb will have collected data corresponding to an integrated luminosity of 8 $fb^{-1}$. Assuming that efficiencies and signal-to-background
ratio will stay unvaried, it is possible to predict an improvement on current limits by
simply scaling them according to the increased statistics. The improvements due to the data collected during Run II should be 
limited (Tab.1).

After the Upgrade phase, with $\sqrt{s}$ = 14 TeV, with a total amount of 50 $fb^{-1}$, one more order of magnitude could be reachable (Tab.1), allowing to confirm or discredit some NP models predictions which still lay between current limits and the SM theory.

As written in the introduction, there is a stimulating field which is still to be investigated, and represents a great chance to test NP theories and SM too. Indeed, the multibody charm decays allow to explore CP \cite{14}, forward-backward \cite{15}-\cite{16} and also T-odd asymmetries \cite{13} fields.
Enhancements of $\mathcal{O}(1\%)$ asymmetries could be measurable in $D^+ \rightarrow \pi^+ \mu \mu$, while $\mathcal{O}(5\%)$ may be detectable with $D^0\rightarrow \pi^+ \pi^− \mu \mu$ (Tab.2).
 
However, these expectations could improve significantly under the upgrade conditions due to the contribution of improvements in the analyses, combination of modes, which might matter more than individual sensitivities, and an offline reconstrucion quality available in a fully software trigger, with an improvement of efficiencies that could reach a factor 3.

\begin{table}
\begin{center}
\begin{tabular}{|c|c|c|c|}\hline
Mode & Run I & Run II & Upgrade \\ \hline
$D^0\rightarrow h h'\mu^+ \mu^-$ & few $10^{-7}$ & fewer $10^{-7}$ & $10^{-8}$ \\ \hline
$D^0\rightarrow \mu^+ \mu^-$ & few $10^{-9}$ & fewer $10^{-9}$ & $10^{-10}$\\ \hline
$D^+\rightarrow \pi^+ \mu^+ \mu^-$ & few $10^{-8}$ & fewer $10^{-8}$ & $10^{-9}$\\ \hline
$D^+_s\rightarrow K^+ \mu^+ \mu^-$ & few $10^{-7}$ & fewer $10^{-7}$ & $10^{-8}$\\ \hline \hline
$\Lambda\rightarrow p \mu\mu$ & few $10^{-7}$ & fewer $10^{-7}$ & $10^{-8}$ \\ \hline 
$D^0\rightarrow e\mu$ & few $10^{-8}$& fewer $10^{-8}$ & $10^{-9}$\\ \hline
$\sigma_{A_{CP}}(D^0	\rightarrow \phi \gamma)$ & $\sim10\%$ & $\sim5\%$ & \\ \hline
\end{tabular}
\label{tab:2}
\caption{Prediction on upper limits on branching fractios,based on Run I results (second column).}
\end{center}
\end{table}

\vspace{-0.3cm}
\begin{table}
\begin{center}
\begin{tabular}{|c|c|c|}\hline
Mode & Run II & Upgrade \\ \hline
$D^+\rightarrow \pi^+ \mu^+ \mu^-$ & $0.6\%$(30000 ev.) & $0.2\%$(300000 events) \\ \hline
$D^0\rightarrow \pi^+ \pi^-\mu^+ \mu^-$ & $3\%$(1500 ev.) & $1\%$(15000 events)\\ \hline
$D^0\rightarrow K^- \pi^+ \mu^+ \mu^-$ & $1\%$(10000 ev.) & $0.3\%$(100000 events)\\ \hline
$D^0\rightarrow K^+ \pi^- \mu^+ \mu^-$ & $40\%$(30 ev.) & $12\%$(300 events)\\ \hline$D^0\rightarrow K^+ K^- \mu^+ \mu^-$ &  $11\%$(150 events) & $4\%$(1500 events) \\ \hline 
\end{tabular}
\label{tab:3}
\caption{Prediction on sensitivities to asymmetries in multi-body rare decays.}
\end{center}
\end{table}

\section{Conclusion}
The results presented are all best world limits.

An update of $D^0\rightarrow \mu^+ \mu^-$ is currently in progress and the results obtained in the search for the Lepton Flavour Violating $D^0\rightarrow e^\pm \mu^\mp$ will become public soon.
Another
measurement, the $D^0\rightarrow K^\mp \pi^\pm \mu^+ \mu^-$ branching fraction should very soon be public. It will serve as a normalisation mode
in the future analyses ssearchinf for NP in $D^0\rightarrow h h' \mu^+ \mu^-$ decays, that are presently on going.

In the next decade, the available datasets will be multiplied by two orders of magnitude. It will make 
more sophisticated measurements, like CP or angular asymmetries, possible.


 
\end{document}

%% file: econfmacros.tex
\def\beq{\begin{equation}}
\def\eeq#1{\label{#1}\end{equation}}
\def\eeqn{\end{equation}}

\def\beqa{\begin{eqnarray}}
\def\eeqa#1{\label{#1}\end{eqnarray}}
\def\eeqan{\end{eqnarray}}

\let\bar=\overbar

\def\Dslash{\not{\hbox{\kern-4pt $D$}}}
\def\dslash{\not{\hbox{\kern-2pt $\del$}}}

\def\msb{{\bar{\ssstyle M \kern -1pt S}}}

